\documentclass[11pt]{article}

\usepackage{graphicx,amsmath,amsfonts,amssymb,dcolumn,amsthm,euscript}

\begin{document}

\title{\textbf{Non-perturbative BRST quantization of Euclidean Yang-Mills theories in Curci-Ferrari gauges}}
\author{\textbf{A.~D.~Pereira$^{1,2,3}$}\thanks{aduarte@if.uff.br}\ , \textbf{R.~F.~Sobreiro$^1$}\thanks{sobreiro@if.uff.br}\ , \textbf{S.~P.~Sorella$^3$}\thanks{silvio.sorella@gmail.com}\\\\
\textit{{\small $^1$UFF $-$ Universidade Federal Fluminense,}}\\
\textit{{\small Instituto de F\'{\i}sica, Campus da Praia Vermelha,}}\\
\textit{{\small Avenida General Milton Tavares de Souza s/n, 24210-346,}}\\
\textit{{\small Niter\'oi, RJ, Brasil}}\and
\textit{{\small $^2$Max Planck Institute for Gravitational Physics, Albert Einstein Institute,}}\\
\textit{{\small Am M\"uhlenberg 1, Potsdam, 14476 Germany}}\and
\textit{{\small $^3$UERJ $-$ Universidade do Estado do Rio de Janeiro,}}\\
\textit{{\small Departamento de F\'isica Te\'orica, Rua S\~ao Francisco Xavier 524,}}\\
\textit{{\small 20550-013, Maracan\~a, Rio de Janeiro, Brasil}}}
\date{}
\maketitle

\begin{abstract}
In this paper we address the issue of the non-perturbative quantization of  Euclidean Yang-Mills theories in the Curci-Ferrari gauge. In particular, we construct  a Refined Gribov-Zwanziger action for this gauge which takes into account the presence of gauge copies as well as the dynamical formation of dimension two condensates. This action enjoys a non-perturbative BRST symmetry recently proposed in \cite{Capri:2015ixa}. Finally, we give attention to the gluon propagator in different space-time dimensions.
\end{abstract}

\section{Introduction}

A fundamental task in theoretical physics is the understanding of non-perturbative aspects of Yang-Mills theories due to the confinement of quarks and gluons, see \cite{Greensite:2011zz} for a general and updated overview. In this regime, the standard well developed perturbation theory is not meaningful and different techniques must be invoked. So far, we have a good toolbox to access the strongly coupled regime with different approaches as lattice simulations, Dyson-Schwinger equations, functional renormalization group methods, holographic techniques, effective models and others, see \cite{Greensite:2011zz,Brambilla:2014jmp}. Despite the impressive progress achieved in the last decade, it seems fair to state that many aspects of the confinement are still to be unraveled.

A long standing problem in the quantization of Yang-Mills theories is the presence of Gribov/gauge copies after the imposition of the gauge fixing condition\footnote{For a pedagogical introduction to the Gribov problem we refer to \cite{Sobreiro:2005ec,Vandersickel:2012tz}.} \cite{Gribov:1977wm}. Within the Faddeev-Popov quantization procedure, these spurious configurations are still being taken into account in the path integral. In particular, a subclass of copies corresponds to zero-modes of the Faddeev-Popov operator making the gauge fixing procedure itself ill-defined, see \cite{Gribov:1977wm,Sobreiro:2005ec,Vandersickel:2012tz}. This feature can be  illustrated in a simple way in the Landau gauge, namely $\partial_{\mu}A^{a}_{\mu}=0$, where we are considering our space-time as a $d$-dimensional Euclidean space with $SU(N)$ gauge group. Performing an infinitesimal gauge transformation over $A^{a}_{\mu}$,

\begin{equation}
\partial_{\mu}A^{a}_{\mu}=0\,\,\mapsto\,\, \partial_{\mu}A'^{a}_{\mu}=\partial_{\mu}\left(A^{a}_{\mu}-D^{ab}_{\mu}(A)\xi^{b}\right)=0\,\,\,\Rightarrow\,\,\, -\partial_{\mu}D^{ab}_{\mu}(A)\xi^b\equiv \EuScript{M}^{ab}_{\mathrm{L}}(A)\xi^b=0
\label{intro1.0}
\end{equation}

\noindent where $\xi^a$ is the infinitesimal parameter of the gauge transformation, $\EuScript{M}^{ab}_{\mathrm{L}}(A)$ is the Faddeev-Popov operator in the Landau gauge and $D^{ab}_{\mu}(A)=\delta^{ab}\partial_{\mu}-gf^{abc}A^{c}_{\mu}$ is the covariant derivative in the adjoint representation of the $SU(N)$ group. From eq.(\ref{intro1.0}), we see that the equivalent configuration $A'^{a}_{\mu}$ satisfies the same condition as $A^{a}_{\mu}$ if $\EuScript{M}^{ab}_{\mathrm{L}}(A)$ develops zero-modes. In \cite{Gribov:1977wm} Gribov showed that the operator $\EuScript{M}^{ab}_{\mathrm{L}}(A)$ develops zero-modes and therefore we have a residual gauge symmetry after the gauge fixing $\partial_{\mu}A^{a}_{\mu}=0$. Gauge copies generated by infinitesimal gauge transformations are called infinitesimal copies. We still have the possibility of generating copies from finite gauge transformations and they do exist indeed \cite{vanBaal:1991zw}.

Already in \cite{Gribov:1977wm}, Gribov proposed a partial solution, in the Landau gauge,  to remove gauge copies from the domain of integration of the path integral by restricting it to the so-called \textit{Gribov region} $\Omega_\mathrm{L}$, which is free of infinitesimal copies. This region is defined as

\begin{equation}
\Omega_{\mathrm{L}} = \{ \; A^a_\mu,  \;  \partial_\mu A^a_\mu =0,  \;    \EuScript{M}^{ab}_{\mathrm{L}}(A) > 0 \; \}    \,, 
\label{intro2.0}
\end{equation}

\noindent and enjoys very important properties: $(i)$ It is bounded in every direction in field space; $(ii)$ it is convex; $(iii)$ it contains the trivial vacuum $A=0$ configuration and; $(iv)$ all gauge orbits cross it at least once. These results were proved in a rigorous fashion in \cite{Dell'Antonio:1991xt} and give a strong support to Gribov's idea to restrict the path integral domain to $\Omega_{\mathrm{L}}$. We should mention, however, that $\Omega_{\mathrm{L}}$ is not free from Gribov copies. Additional copies still exist inside $\Omega_{\mathrm{L}}$ \cite{vanBaal:1991zw}. Nevertheless, it is possible to define a subset $\Lambda$ of $\Omega_{\mathrm{L}}$ which is fully free from gauge copies. The region $\Lambda$ is known as the \textit{fundamental modular region} (FMR). Though, so far, the practical implementation of the restriction of the domain of integration in the path integral has been worked out only for the Gribov region $\Omega_{\mathrm{L}}$. 

Formally, Gribov's proposal is written as

\begin{equation}
\EuScript{Z}=\int_{\Omega_{\mathrm{L}}} \left[\EuScript{D}\mathbf{\Phi}\right]\mathrm{e}^{-\left( S_{\mathrm{YM}}+S_{\mathrm{gf}}\right) }\,.
\label{intro3.0}
\end{equation}

\noindent In his original paper, Gribov implemented this restriction in the Landau gauge up to leading order in perturbation theory. Subsequently,  this computation was generalized to all orders by Zwanziger in \cite{Zwanziger:1989mf}. Although their methods are different, it turns out that they lead to equivalent results \cite{Capri:2012wx}. The result worked out by Zwanziger shows that the restriction to $\Omega_{\mathrm{L}}$ can be effectively implemented by the addition of a non-local term to the standard gauge fixed Yang-Mills action and of a vacuum term giving rise to the so called Gribov-Zwanziger action,  

\begin{equation}
\int_{\Omega_{\mathrm{L}}} \left[\EuScript{D}\mathbf{\Phi}\right]\mathrm{e}^{-\left( S_{\mathrm{YM}}+S_{\mathrm{gf}}\right) } = \int \left[\EuScript{D}\mathbf{\Phi}\right]\mathrm{e}^{-    {S}^{L}_{\mathrm{GZ}}  }   \;, 
\label{intro4.0}
\end{equation}

\noindent with
 
\begin{equation}
{S}^{L}_{\mathrm{GZ}}=   S_{\mathrm{YM}}+S_{\mathrm{gf}}    +\gamma^4 H_{L}(A) - dV\gamma^4(N^2-1)\,, 
\label{intro5.0}
\end{equation}
where $(S_{\mathrm{YM}}, S_{\mathrm{gf}})$ denote, respectively,  the Yang-Mills action and the Faddeev-Popov term corresponding to the Landau gauge-fixing, namely 
\begin{equation} 
S_{\mathrm{YM}} = \frac{1}{4} \int d^dx \;  F^a_{\mu\nu} F^a_{\mu \nu} \;, \label{ymact}
\end{equation}
\begin{equation}
S_{\mathrm{gf}} =\int d^dx\left(b^{a}\partial_{\mu}A^{a}_{\mu}+\bar{c}^{a}\partial_{\mu}D^{ab}_{\mu}c^b\right) \;, \label{lgfa} 
\end{equation}

\noindent and 

\begin{equation}
H_{L}(A)=g^2\int d^dxd^dy~f^{abc}A^{b}_{\mu}(x)\left[\EuScript{M}^{-1}_{\mathrm{L}}(A)\right]^{ad}(x,y)f^{dec}A^{e}_{\mu}(y)\,, 
\label{intro6.0}
\end{equation}

\noindent is known as the \textit{horizon function}. The quantity  $V$ in expression \eqref{intro5.0} represents the Euclidean volume in $d$-dimensional space-time, while $\gamma$ is the so-called \textit{Gribov parameter}, a mass parameter which naturally emerges from the restriction to $\Omega_{\mathrm{L}}$. This parameter, however, is not free, being determined in a self consistent way through the gap equation (or horizon condition)

\begin{equation}
\langle H_{L}(A) \rangle = dV(N^2-1)\,,
\label{intro7.0}
\end{equation}

\noindent where expectation values are taken with respect to the modified measure of expression  (\ref{intro4.0}). It is apparent  from the presence of the inverse of the Faddeev-Popov operator $\EuScript{M}^{-1}_{\mathrm{L}}$ that the Gribov-Zwanziger action is non-local. Notably,  it can be cast in local form by the introduction of a suitable set of auxiliary fields, namely, a pair of commuting ones $(\bar{\varphi}^{ab}_{\mu},\varphi^{ab}_{\mu})$ and another pair of anti-commuting fields $(\bar{\omega}^{ab}_{\mu},\omega^{ab}_{\mu})$. The expression for the local Gribov-Zwanziger action is given by

\begin{eqnarray}
S^{L}_{\mathrm{GZ}} &=& S_{\mathrm{YM} } +S_{\mathrm{gf}}-\int d^dx\left(\bar{\varphi}^{ac}_{\mu}\EuScript{M}^{ab}_{\mathrm{L}}(A)\varphi^{bc}_{\mu}-\bar{\omega}^{ac}_{\mu}\EuScript{M}^{ab}_{\mathrm{L}}(A)\omega^{bc}_{\mu} + gf^{adb}(\partial_{\nu}\bar{\omega}^{ac}_{\mu})(D^{de}_{\nu}c^{e})\varphi^{bc}_{\mu}\right) \nonumber \\
&+& \gamma^2\int d^dx~gf^{abc}A^{a}_{\mu}(\varphi+\bar{\varphi})^{bc}_{\mu}-d\gamma^4V(N^2-1)\,,
\label{intro8.0}
\end{eqnarray}

\noindent and is easy to check that, upon integration over the auxiliary fields $(\bar{\varphi}^{ab}_{\mu},\varphi^{ab}_{\mu},  \bar{\omega}^{ab}_{\mu},\omega^{ab}_{\mu})$,  we re-obtain expression (\ref{intro5.0}). Notice also that the fields $(\bar{\varphi}^{ab}_{\mu},\varphi^{ab}_{\mu},  \bar{\omega}^{ab}_{\mu},\omega^{ab}_{\mu})$ carry both Lorentz and color indices in the adjoint representation of the gauge group, {\it i.e.} $(a,b)=1....(N^2-1)$. In this local picture, the gap equation (\ref{intro7.0}) is expressed as

\begin{equation}
\frac{\partial\mathcal{E}_0}{\partial\gamma^2}=0\,,
\label{intro9.0}
\end{equation}
where $\mathcal{E}_0$ stands for the vacuum energy of the theory, {\it i.e.}
\begin{equation}
\mathrm{e}^{-V\mathcal{E}_0}=\int \left[\EuScript{D}\Phi\right]\mathrm{e}^{-S^{L}_{\mathrm{GZ}}}\,,
\label{intro9.1}
\end{equation}
with $\Phi$ the complete set of fields. 

Remarkably, the Gribov-Zwanziger action (\ref{intro8.0}) is renormalizable to all orders in perturbation theory \cite{Vandersickel:2012tz,Zwanziger:1992qr}. Thus, the action  (\ref{intro8.0})  provides a local and renormalizable framework to deal with the existence of (infinitesimal) copies. This action also displays a very interesting feature: It breaks the BRST symmetry explicitly, although in a soft way. In particular,

\begin{equation}
sS^{L}_{\mathrm{GZ}} = g\gamma^2\int d^dx~f^{abc}\left[-D^{ad}_{\mu}c^d(\varphi+\bar{\varphi})^{bc}_{\mu}+A^{a}_{\mu}\omega^{bc}_{\mu}\right]\,,
\label{intro10.0}
\end{equation}

\noindent whereby we can see that the explicit breaking is soften by the Gribov parameter $\gamma$. Solving the gap equation (\ref{intro9.0}) to leading order, it is possible to show $\gamma^2\propto\exp{\left(-1/g^2\right)}$. This exhibits the non-perturbative nature of $\gamma$ since, in the deep ultraviolet region, it implies that $\gamma^2\rightarrow 0$. Also, from eq.(\ref{intro10.0}), the BRST invariance is recovered in the UV. Nevertheless, in the IR the breaking is present. This BRST soft breaking is one of the outstanding points of the Gribov-Zwanziger scenario and is debated up to date, see \cite{Dudal:2009xh,Sorella:2009vt,Baulieu:2008fy,Capri:2010hb,Dudal:2012sb,Dudal:2014rxa,Pereira:2013aza,Pereira:2014apa,Capri:2014bsa,Tissier:2010ts,Serreau:2012cg,Serreau:2013ila,Serreau:2015yna,Lavrov:2013boa,Moshin:2015gsa,Schaden:2014bea,Cucchieri:2014via,Dudal:2010hj}.

Tracing back to the foundations of Gribov and Zwanziger ideas to remove gauge copies from the domain of integration of the path integral,  it is rather clear that  the Hermitian nature of the Faddeev-Popov operator $\EuScript{M}_{L}(A)$ in the Landau gauge plays a pivotal role. In particular, the very definition of the Gribov region $\Omega_{\mathrm{L}}$ relies on the positivity of $\EuScript{M}_{L}(A)$, a meaningful concept due to the real spectrum of such operator. Nevertheless, hermiticity of the Faddeev-Popov operator is generally lost outside of the Landau gauge. This is the case, for instance, of the linear covariant gauges. Therefore, a very natural question arises, since the Gribov problem is not a particular feature of Landau gauge, but of all gauge conditions that are continuous in field space \cite{Singer:1978dk}: How to construct a consistent resolution of this problem in different gauges? 

This requires the need for strategies different than the one described above. Also, in perturbative gauge theories, BRST symmetry plays a prominent role in the proof of gauge independence of physical operators. Since in the Gribov-Zwanziger setting in the Landau gauge BRST is broken, it is rather natural to expect this will pose some difficulties in the proof of gauge-independence as soon as we move away from Landau gauge. It is worth mentioning that the consruction of the Gribov-Zwanziger action following the aforementioned method to maximal Abelian and Coulomb gauges is viable due to the hermiticity of their Faddeev-Popov operators, see \cite{Pereira:2013aza,Gongyo:2013rua,Capri:2010an,Capri:2013vka,Capri:2008vk}. Different frameworks to handle the Gribov problem in a one-parameter family of Landau gauges were proposed in \cite{Serreau:2012cg,Serreau:2013ila,Serreau:2015yna}. Though, a soft BRST breaking is still present.  

Recently, particular attention was devoted to the linear covariant gauges in the Gribov-Zwanziger context\footnote{See \cite{Siringo:2014lva,Aguilar:2015nqa,Huber:2015ria,Bicudo:2015rma} for developments outside the Gribov-Zwanziger set up.} \cite{Capri:2015ixa,Capri:2015pja,Capri:2015nzw,Capri:2016aqq}, see also \cite{Sobreiro:2005vn} for the very first attempt. These gauges bring two challenging features for the Gribov-Zwanziger set up: First, the gauge condition is given by

\begin{equation}
\partial_{\mu}A^{a}_{\mu}=\alpha b^a\,,
\label{11.0}
\end{equation} 
with $\alpha$ a non-negative gauge parameter and $b^a$ a fixed field configuration. This entails a non-Hermitian Faddeev-Popov operator. Second, the presence of the gauge parameter $\alpha$ allows us to check the gauge independence of physical quantities in a very explicit way.  

Dealing with this problem has enabled us to introduce a  non-perturbative BRST symmetry \textit{i.e.} a set of transformations corresponding to a non-perturbative generalization of the standard BRST transformations which result in an exact symmetry of the Gribov-Zwanziger action \cite{Capri:2015ixa}. Furthermore, this non-perturbative symmetry turns out to be  generated by a nilpotent operator, a feature which preserves the important concept of the BRST cohomology. Remarkably, this framework allows for the construction of a Gribov-Zwanziger action in linear covariant gauges which, due to its exact non-perturbative BRST symmetry, enjoys the important feature that the correlation functions of quantities which are invariant under the new non-perturbative  BRST symmetry are in fact independent from the gauge parameter $\alpha$.  

In this paper, we address the quantization of a class of non-linear gauges, known as Curci-Ferrari gauges \cite{Baulieu:1981sb,Delduc:1989uc}, generalizing the non-perturbative BRST introduced in the class of the linear covariant gauges  \cite{Capri:2015ixa,Capri:2015pja,Capri:2015nzw}. As we shall discuss, the non-linearity of these gauges brings novel effects as the formation of ghosts and gluon-ghosts condensates. This topic was already investigated in \cite{Kondo:2001nq,Dudal:2003gu,Dudal:2002ye}. In this paper we will comment on these effects in light of the new non-perturbative BRST symmetry. We emphasize that recent studies on the same issue were done in \cite{Serreau:2013ila,Serreau:2015yna}.

Though, before going any further, it is worth spending a few words on the important and intensively investigated issue of the understanding of the behavior of the gluon propagator in the non-perturbative IR regime \cite{Maas:2011se}. It is widely accepted  that the IR analysis of the gluon two-point correlation function brings us quantitative information about gluon confinement. In particular, very recent lattice simulations in the Landau gauge have shown an IR suppressed, positivity violating gluon propagator which attains a finite non-vanishing value at zero-momentum in $d=4$ \cite{Cucchieri:2007rg,Cucchieri:2008fc,Maas:2008ri,Cucchieri:2011ig}. The violation of positivity hinders a K\"all\'en-Lehmann representation of such propagator which makes impossible the interpretation of gluons as stable particles  of the physical spectrum. This positivity  violation of the gluon propagator occurs at the confinement scale and is interpreted as a strong  signal of gluon confinement. Also, due to the finite value of the form factor at zero-momentum, this propagator is referred to as the  \textit{decoupling/massive} solution. A similar behavior for the Landau gluon propagator is found in $d=3$, while in $d=2$  the gluon propagator attains a vanishing value at zero-momentum, see \cite{Cucchieri:2011ig}.  In this latter case, the propagator is known as being of the \textit{scaling} type. 

Working out the two-point gluon correlation function  from expression \eqref{intro8.0}, it turns out that the Gribov-Zwaniger framework gives rise to a propagator which is 
of the scaling type for $d=2,3,4$. However, in \cite{Dudal:2007cw,Dudal:2008sp}, it was noted that the Gribov-Zwanziger action suffers from IR instabilities giving rise to the formation of dimension-two condensates. In particular, albeit introduced to cast the framework in a local fashion, the auxiliary fields $(\bar{\varphi},\varphi,\bar{\omega},\omega)$ develop their own dynamics which results in the formation of dimension-two condensates in both $d=3,4$ \cite{Dudal:2008sp,Dudal:2008rm,Dudal:2008xd}. These condensates modify the behavior of the gluon propagator, turning it from scaling  to a decoupling type. Moreover, as discussed in \cite{Dudal:2008sp,Dudal:2008rm,Dudal:2008xd}, in $d=2$ the formation of the aforementioned condensates cannot occur due to the presence of infrared singularities, a fact in agreement with the observed scaling behavior for the gluon propagator. The introduction of such condensates in $d=3,4$ gives rise to the so-called Refined Gribov-Zwanziger action \cite{Dudal:2008sp,Dudal:2008rm,Dudal:2008xd} which predicts a decoupling type gluon propagator. Moreover, in $d=2$, infrared singularities prevent the formation of the condensates. As a consequence, in $d=2$, the  Gribov-Zwanziger action does not suffer from refinement, generating a scaling type gluon propagator. These results are in good qualitative agreement with recent lattice numerical simulations \cite{Cucchieri:2011ig,Dudal:2012hb}.

A non-trivial fact is that these results on the IR behavior of the gluon propagator are not peculiar to the Landau gauge, but also displayed by the Gribov-Zwanziger approach to linear covariant, maximal Abelian and Coulomb gauges, \cite{Capri:2015nzw,Capri:2015pfa,Guimaraes:2015bra}, a feature which suggests the possible existence of a general pattern for the gluon two-point correlation function. In this paper we show how this extends to Curci-Ferrari gauges as well.

The paper is organized as follows: Sect.~2 is devoted to a brief historical overview of the relevance of the Curci-Ferrari gauges. In Sect.~3 we provide a review of the non-perturbative BRST quantization in the linear covariant gauges and set the key tools of the framework. In Sect.~4, the Gribov problem in the Curci-Ferrari gauges is discussed. In particular, we shall be able to show that, by means of a suitable redefinition of the Lagrange multiplier $b^a$, it  can addressed in a similar way to the case of the linear covariant gauges, allowing us to use the techniques introduced in Sect.~3. Sect.~5  is devoted to the construction of the Gribov-Zwanziger action for the Curci-Ferrari gauges, while in Sect.~6 we present its refined  version. We devote Sect.~7 to the tree-level gluon propagator computed with the Refined Gribov-Zwanziger action.  In Sect.~8, a local form of the Refined Gribov-Zwanziger action is obtained, providing thus a local and non-perturbative BRST invariant framework for the Curci-Ferrari gauges. Finally, we present our conclusions and perspectives.

\section{Usefulness of the Curci-Ferrari model}

In this paper the Gribov problem is addressed in the Curci-Ferrari gauges. One might ask if there are reasons to study such problem in a rather non-trivial gauge such the Curci-Ferrari case beyond an intrinsic interest on the Gribov problem itself. It is worth emphasizing, though, that different investigations of non-perturbative effects generated in Yang-Mills theories quantized in Curci-Ferrari gauges have witnessed great interest since many years. 

In particular, the study of the condensation of dimension-two operators was closely analysed in \cite{Kondo:2001nq,Dudal:2003gu,Dudal:2002ye,Dudal:2003dp,Alkofer:2003jr}, bringing novel non-perturbative modifications to the gluon and ghost propagators. Moreover, investigations on the lattice formulation of the BRST quantization of Yang-Mills theory have already relied on the use of  the Curci-Ferrari model, see for example \cite{Kalloniatis:2005if,Ghiotti:2006pm,vonSmekal:2008en}. In fact, it turns out that the inclusion of the so-called Curci-Ferrari mass provides a regularization for the well-known Neuberger problem, see \cite{Neuberger:1986xz}. More recently, the Curci-Ferrari action was used to construct a non-perturbative miodel  for Yang-Mills theory whose results are in agreement with the decoupling/massive solution for the gluon propagator, without making use of the Higgs mechanism, \cite{Kondo:2012ac,Kondo:2012ri}. 

Let us also mention that the issue of the Gribov copies in Yang-Mills theory quantized in Curci-Ferrari gauges was recently addressed in \cite{Serreau:2013ila,Serreau:2015yna}, where it was shown that they  affect significantly the infrared regime of the theory. In particular, in the present  paper, such effects are studied within the context of the recently proposed non-perturbative BRST symmetry \cite{Capri:2015ixa} which emerges from the elimination of the Gribov copies. 

These considerations provide  a good motivation for the present investigation, while giving an overview of the efforts which are currently done, from both analytical and numerical sides, to access the non-perturbative regime of Yang-Mills theory quantized in gauges different from the Landau gauge.   

\section{An overview of the non-perturbative BRST quantization}

In this section, we briefly review the main aspects of the non-perturbative BRST quantization introduced in \cite{Capri:2015ixa,Capri:2015nzw}. To begin with, we construct the non-perturbative BRST symmetry from a reformulation of the Gribov-Zwanziger action \eqref{intro5.0} in terms of the transverse gauge invariant field

\begin{equation}
A_{\mu }^{h} =\left( \delta _{\mu \nu }-\frac{\partial _{\mu }\partial
_{\nu }}{\partial ^{2}}\right) \left( A_{\nu }-ig\left[ \frac{1}{\partial
^{2}}\partial A,A_{\nu }\right] +\frac{ig}{2}\left[ \frac{1}{\partial ^{2}}%
\partial A,\partial _{\nu }\frac{1}{\partial ^{2}}\partial A\right] \right)
+O(A^{3})\,,
\label{overview1.0}
\end{equation}
obtained from the minimization with respect to $U$ of the functional

\begin{equation}
\mathcal{A}^{2}_{\mathrm{min}}=\underset{\left\{U\right\}}{\mathrm{min}}\,\mathrm{Tr\,}\int d^dx\,A^{U}_{\mu}A^{U}_{\mu}\,,
\label{overview2.0}
\end{equation}
with 

\begin{equation}
A^{U}_{\mu}=U^{\dagger}A_{\mu}U+\frac{i}{g}U^{\dagger}\partial_{\mu}U\,,
\label{overview3.0}
\end{equation}
where $U$ is a $SU(N)$ matrix. Working out the minimization process, for  $\mathcal{A}^{2}_{\mathrm{min}}$ one gets the highly non-local, albeit gauge invariant, expression 
\begin{equation}
\mathcal{A}^{2}_{\mathrm{min}}= \mathrm{Tr} \int d^dx\,A^{h}_{\mu}A^{h}_{\mu}\,.
\label{a2min}
\end{equation}

Our conventions are such that $A_{\mu}=A^{a}_{\mu}T^{a}$, with $T^a$ denoting the hermitian $SU(N)$ generators satisfying the algebra $[T^a,T^b]=if^{abc}T^c$. For details on the derivation of \eqref{overview1.0} we refer to \textit{e.g.} \cite{Capri:2015ixa}. From expression \eqref{overview1.0} it is clear that the formal power series starts with $A_{\mu}$ itself and then all terms contain at least one power of $\partial A$. With this in mind, we can rewrite the horizon function \eqref{intro6.0} as 

\begin{equation}
H(A)=H(A^h)-\int d^dx\, d^dy~R^a(x,y)(\partial_{\mu} A^a_{\mu})_y\,,
\label{overview4.0}
\end{equation}
where the explicit form of $R^a$ is not relevant for our purposes. The Gribov-Zwanziger action in Landau gauge is then rewritten as 

\begin{equation}
\tilde{S}_{\mathrm{GZ}}=S_{\mathrm{YM}}+\int d^dx\left(b^{h,a}\partial_{\mu}A^{a}_{\mu}+\bar{c}^{a}\partial_{\mu}D^{ab}_{\mu}c^b\right)+\gamma^4H(A^h)\,,
\label{overview5.0}
\end{equation}
where $b^{h,a}$ stands for the redefined Lagrange multiplier

\begin{equation}
b^{h,a}=b^a-\gamma^4R(A)\,.
\label{overview6.0}
\end{equation}
Notice that expression \eqref{overview6.0} corresponds to a field redefinition with unit Jacobian, as it is easily checked.  Again, we introduce the Zwanziger auxiliary fields $(\bar{\varphi},\varphi,\bar{\omega},\omega)$ and rewrite the Gribov-Zwanziger action as

\begin{eqnarray}
S_{\mathrm{GZ}}&=& S_{\mathrm{YM}}+\int d^dx\left(b^{h,a}\partial_{\mu}A^{a}_{\mu}+\bar{c}^{a}\partial_{\mu}D^{ab}_{\mu}c^{b}\right)\nonumber\\
&-&\int d^dx\left(\bar{\varphi}^{ac}_{\mu}\left[\EuScript{M}(A^h)\right]^{ab}\varphi^{bc}_{\mu}-\bar{\omega}^{ac}_{\mu}\left[\EuScript{M}(A^h)\right]^{ab}\omega^{bc}_{\mu}+g\gamma^2f^{abc}A^{h,a}_{\mu}(\varphi+\bar{\varphi})^{bc}_{\mu}\right)\,.
\label{overview7.0}
\end{eqnarray}
Differently from the standard Gribov-Zwanziger formulation in the Landau gauge \eqref{intro8.0}, the action \eqref{overview7.0} is not local after the introduction of the auxiliary fields. The reason is that, when written in terms of $A^h$, the horizon function displays two sources of non-localities: First, the standard one coming from the presence of the inverse of $\EuScript{M}$. The second type of non-locality comes from $A^h$ itself, which so far was written as a formal power series as in \eqref{overview1.0}. Nevertheless, action \eqref{overview7.0} enjoys a \textit{non-perturbative} BRST symmetry given by

\begin{align}
s_{\gamma^2}A^{a}_{\mu}&=-D^{ab}_{\mu}c^b\,,     &&s_{\gamma^2}c^a=\frac{g}{2}f^{abc}c^bc^c\,, \nonumber\\
s_{\gamma^2}\bar{c}^a&=b^{h,a}\,,     &&s_{\gamma^2}b^{h,a}=0\,, \nonumber\\
s_{\gamma^2}\varphi^{ab}_{\mu}&=\omega^{ab}_{\mu}\,,   &&s_{\gamma^2}\omega^{ab}_{\mu}=0\,, \nonumber\\
s_{\gamma^2}\bar{\omega}^{ab}_{\mu}&=\bar{\varphi}^{ab}_{\mu}-\gamma^2gf^{cdb}\int d^dy~A^{h,c}_{\mu}(y)\left[\EuScript{M}^{-1}(A^h)\right]^{da}_{yx}\,,         &&s_{\gamma^2}\bar{\varphi}^{ab}_{\mu}=0\,, 
\label{overview8.0}
\end{align}
with 

\begin{equation}
s_{\gamma^2}S_{\mathrm{GZ}}=0\,\,\,\,\,\,\mathrm{and}\,\,\,\,\, s^{2}_{\gamma^2}=0\,.
\label{oveview9.0}
\end{equation}
We emphasize that the gauge invariance of $A^h$ automatically implies

\begin{equation}
sA^{h,a}_{\mu}=0 \,\,\, \Rightarrow \,\,\, s_{\gamma^2}A^{h,a}_{\mu}=0\,,
\label{overview10.0}
\end{equation}
because $sA_{\mu}=s_{\gamma2}A_{\mu}$. As discussed in \cite{Capri:2015ixa}, the transformations generated by $s_{\gamma^2}$ are such that when $\gamma\rightarrow 0$, $s_{\gamma^2}\rightarrow s$. Therefore, in the perturbative regime where the Gribov parameter can be set to zero, we recover the standard BRST transformations. Due to the explicit presence of the non-perturbative Gribov parameter $\gamma^2$ in eqs.\eqref{overview8.0}, the nilpotent  operator $s_{\gamma^2}$ can be naturally seen as a non-perturbative extension of the standard BRST operator $s$.

Therefore, when written in terms of the variable $A^h$, the Gribov-Zwanziger action displays an exact and nilpotent non-perturbative BRST symmetry given by \eqref{overview8.0}. Also, when employing $A^h$, the Gribov-Zwanziger action is non-local even after the introduction of the Zwanziger auxiliary fields as well as the BRST transformations. Nevertheless, recently, a full localization of the entire set up has been constructed, as  presented in \cite{Capri:2016aqq}.

With \eqref{overview8.0} at our disposal, a non-perturbative BRST quantization was proposed in \cite{Capri:2015ixa,Capri:2015nzw} for the linear covariant gauges. The resulting Gribov-Zwanziger action in linear covariant gauges is

\begin{eqnarray}
S^{\mathrm{LCG}}_{\mathrm{GZ}}&=& S_{\mathrm{YM}}+s_{\gamma^2}\int d^dx~\bar{c}^{a}\left(\partial_{\mu}A^{a}_{\mu}-\frac{\alpha}{2}b^{h,a}\right)\nonumber\\
&-&\int d^dx\left(\bar{\varphi}^{ac}_{\mu}\left[\EuScript{M}(A^h)\right]^{ab}\varphi^{bc}_{\mu}-\bar{\omega}^{ac}_{\mu}\left[\EuScript{M}(A^h)\right]^{ab}\omega^{bc}_{\mu}+g\gamma^2f^{abc}A^{h,a}_{\mu}(\varphi+\bar{\varphi})^{bc}_{\mu}\right)\nonumber\\
&=&  S_{\mathrm{YM}} + \int d^dx\left(b^{h,a}\left(\partial_{\mu}A^{a}_{\mu}-\frac{\alpha}{2}b^{h,a}\right)+\bar{c}^{a}\partial_{\mu}D^{ab}_{\mu}c^{b}\right)\nonumber\\
&-&\int d^dx\left(\bar{\varphi}^{ac}_{\mu}\left[\EuScript{M}(A^h)\right]^{ab}\varphi^{bc}_{\mu}-\bar{\omega}^{ac}_{\mu}\left[\EuScript{M}(A^h)\right]^{ab}\omega^{bc}_{\mu}+g\gamma^2f^{abc}A^{h,a}_{\mu}(\varphi+\bar{\varphi})^{bc}_{\mu}\right)\,.
\label{overview11.0}
\end{eqnarray}
Action \eqref{overview11.0} is manifestly invariant under non-perturbative BRST transformations.

An important aspect of the action \eqref{overview11.0} is that it is possible to show that it restricts the domain of integration in the path integral to a region which is free from a large set of Gribov copies  \cite{Capri:2015ixa,Capri:2015nzw}. It is worth mentioning here that, besides Yang-Mills theories, other models have been investigated within the approaches outlined by Gribov and Zwanziger. Let us quote the case of 2d gravity and string theory \cite{Baulieu:2012ay}, where the powerful and deep knowledge of the topology of two-dimensional Riemann surfaces has allowed to  see explicitly the removal of all Gribov redundancies by the restriction of the path integral to a particular domain. For instance, in the case of the torus, an explicit check of the positivity of the eigenvalues of the Faddeev-Popov operator when restricted to the analogous of the Gribov region has been worked out in \cite{Baulieu:2012ay}.  \\\\Coming back to gauge theories, a non-perturbative BRST quantization leads thus to a Gribov-Zwanziger action in linear covariant gauges. An immediate consequence of the manifest non-perturbative BRST invariance is the  independence from the gauge parameter $\alpha$ of the gap equation \cite{Capri:2015ixa,Capri:2015nzw},

\begin{equation}
\langle H(A^h)\rangle=dV(N^2-1)\,.  
\label{overview12.0}
\end{equation}
This equation determines the Gribov parameter $\gamma$ and its gauge invariance ensures that $\gamma$ itself is independent of $\alpha$, namely, it is a physical parameter of the theory. We remind in fact that the parameter $\gamma$ enters in an explicit way the correlation functions of gauge invariant operators as reported, for instance, in the evaluation of the spectrum of the glueballs in the Landau gauge \cite{Dudal:2010cd,Dudal:2013wja}. The $\alpha$-independence of $\gamma$ is thus an important outcome of the consistency of the non-perturbative BRST set up developed in \cite{Capri:2015ixa,Capri:2015nzw}.   

\section{Establishing the Gribov problem in Curci-Ferrari gauges} \label{conv}

In \cite{Baulieu:1981sb,Delduc:1989uc} a family  of non-linear gauges containing only one gauge parameter was introduced. Quite often, these gauges are called Curci-Ferrari gauges because the Lagrangian is exactly the same introduced in \cite{Curci:1976bt,Curci:1976kh} by Curci and Ferrari. There, however, a mass term for the gluons is introduced to discuss massive Yang-Mills theories. Inhere, we will deal with the massless case.  

\subsection{Conventions and standard BRST quantization}

The gauge fixed Yang-Mills action in Curci-Ferrari gauges in $d$-dimensional Euclidean space-time  is given by

\begin{eqnarray}
S_{\mathrm{FP}}&=& S_{\mathrm{YM}}+s\int d^dx~\bar{c}^{a}\left[\partial_{\mu}A^{a}_{\mu}-\frac{\alpha}{2}\left(b^a-\frac{g}{2}f^{abc}\bar{c}^{b}c^{c}\right)\right]\nonumber\\
&=&S_{\mathrm{YM}}+\int d^dx\left[b^{a}\partial_{\mu}A^{a}_{\mu}+\bar{c}^{a}\partial_{\mu}D^{ab}_{\mu}(A)c^{b}-\frac{\alpha}{2}b^{a}b^{a}+\frac{\alpha}{2}gf^{abc}b^{a}\bar{c}^{b}c^{c}\right.\nonumber\\
&+&\left.\frac{\alpha}{8}g^{2}f^{abc}f^{cde}\bar{c}^{a}\bar{c}^{b}c^{d}c^{e}\right]\,.
\label{c1}
\end{eqnarray}
\noindent This action is manifestly invariant under the standard BRST transformations,

\begin{eqnarray}
sA^{a}_{\mu}&=&-D^{ab}_{\mu}c^{b}\nonumber\\
sc^{a}&=&\frac{g}{2}f^{abc}c^{b}c^{c}\nonumber\\
s\bar{c}^{a}&=&b^{a}\nonumber\\
sb^{a}&=&0\,,
\label{c2}
\end{eqnarray}
and is renormalizable to all orders in perturbation theory \cite{Delduc:1989uc}. It is worth mentioning that action \eqref{c1} contains an interaction term between Faddeev-Popov ghosts and the auxiliary field $b$ and a quartic interaction of ghosts. As we shall see, the presence of such terms is responsible to drive different dynamical effects with respect to linear covariant gauges. In particular, the equations of motion of the auxiliary field $b$ and of the  the anti-ghost $\bar{c}$ do not correspond anymore to Ward identities,  due to the non-linear character of this gauge. We remind that, in the case of the linear covariant gauges, these equations do correspond to Ward identities which play an important role in the proof of the renormalizability. 

On the other hand, action \eqref{c1} enjoys another global symmetry besides BRST which will generate a Ward identity that plays a role analogous to that of the anti-ghost equation in linear gauges. This symmetry is known as the $SL(2,\mathbb{R})$ symmetry\footnote{We refer to App.~A for more details on the $SL(2,\mathbb{R})$ algebra.} and its associated Ward identity, together with the Slavnov-Taylor identity, guarantees the all order proof of renormalizability of such gauge \cite{Delduc:1989uc}. The $SL(2,\mathbb{R})$ symmetry is defined by the following set of transformations:

\begin{eqnarray}
\delta\bar{c}^{a}&=&c^a\nonumber\\
\delta b^{a}&=&\frac{g}{2}f^{abc}c^{b}c^{c}\nonumber\\
\delta A^{a}_{\mu}&=&\delta c^{a} = 0\,,
\label{c3}
\end{eqnarray}
and 
\begin{equation}
\delta S_{\mathrm{FP}}=0 \;.   \label{ds}
\end{equation}

\noindent An useful property is that the $SL(2,\mathbb{R})$ operator $\delta$ commutes with the BRST operator $s$ \textit{i.e.} $[s,\delta]=0$.

\subsection{Construction of a copies equation}

As shown in the example of eq.\eqref{intro1.0}, given a gauge condition\footnote{For example, $F[A]=\partial_\mu A_\mu=0$ in the Landau gauge, while  $F[A]=\partial_\mu A_\mu - \alpha b=0$ in the linear covariant gauges.} $F[A]=0$, we can characterize the existence of Gribov copies by performing a gauge transformation over $F[A]=0$ and looking for solutions of the resulting equation - the copies equation. Nevertheless, in the case of Curci-Ferrari gauges, due to its non-linear character, is not clear how to read off from the action \eqref{c1} the gauge condition $F[A]=0$, with $F$ being  a functional of the gauge field. As it is immediately checked, in the cases of the Landau and linear covariant gauges, the gauge fixing condition, {\it i.e.} $F[A]=0$, is expressed through the equation of motion of the Lagrange multiplier field $b^a$. However, looking at this equation in the case of the Curci-Ferrari gauge, one gets  
\begin{equation}
\frac{\delta S_{\mathrm{FP}}}{\delta b^a}=\partial_{\mu}A^{a}_{\mu}-\alpha b^a + \frac{\alpha}{2}g f^{abc}b^{a}\bar{c}^{b}c^c\,.
\label{cce1}
\end{equation}
One sees thus that, due to the presence of the ghost term $\frac{\alpha}{2}g f^{abc}b^{a}\bar{c}^{b}c^c$, this equation cannot be interpreted as a genuine gauge fixing condition $F[A]=0$. 

On the other hand, it is possible to cast the Curci-Ferrari gauges in a form similar to that  of the linear covariant gauges by means of a  suitable shift on the $b$ field, \cite{Baulieu:1981sb}. To that purpose, we perform the following shift in the path integral 

\begin{equation}
b^a\,\,\longrightarrow\,\, b'^a=b^a-\frac{g}{2}f^{abc}\bar{c}^bc^c\,,
\label{cce2}
\end{equation}
which entails a trivial Jacobian. The Yang-Mills action in Curci-Ferrari gauges is then rewritten as

\begin{equation}
S_{\mathrm{FP}} = S_{\mathrm{YM}}+s\int d^4x~\bar{c}^{a}\left(\partial_{\mu}A^{a}_{\mu}-\frac{\alpha}{2}b'^a\right)\,,
\label{cce3}
\end{equation}

\noindent which looks the same expression as in the linear covariant gauges. However, the difference between the two cases arises from the fact that the corresponding  BRST transformations will now also change. Nevertheless, we can still exploit the similarity between these gauges at the formal level and keep in mind the different roles played by $b$ and $b'$. So, as a gauge-fixing condition,  we express the Curci-Ferrari gauges as

\begin{equation}
\partial_{\mu}A^{a}_{\mu}=\alpha b'^a\,.
\label{cce4}
\end{equation}
We can treat \eqref{cce4} as our desired $F[A]=0$ equation. Since it is formally identical to the gauge-fixing equation for linear covariant gauges, we can immediately conclude that their solutions are formally the same. As a consequence, the framework contructed in \cite{Capri:2015ixa} to deal with the Gribov problem in linear covariant gauges can be employed as well in the case of  the Curci-Ferrari gauges. This is precisely the subject of the next section.

The shifted BRST transformations are expressed as

\begin{eqnarray}
sA^{a}_{\mu}&=&-D^{ab}_{\mu}c^{b}\nonumber\\
sc^{a}&=&\frac{g}{2}f^{abc}c^{b}c^{c}\nonumber\\
s\bar{c}^{a}&=&b'^a+\frac{g}{2}f^{abc}\bar{c}^{b}c^{c}\nonumber\\
sb'^a&=&-\frac{g}{2}f^{abc}b'^bc^c+\frac{g^{2}}{8}f^{abc}f^{cde}\bar{c}^{b}c^{d}c^{e}\,.
\label{cce5}
\end{eqnarray}
Explicitly, the Faddeev-Popov action in terms of the field $b'$ is given by

\begin{equation}
S_{\mathrm{FP}}=S_{\mathrm{YM}}+\int d^dx\left[b'^a\partial_{\mu}A^{a}_{\mu}+\frac{1}{2}\bar{c}^{a}(\partial_{\mu}D^{ab}_{\mu}+D^{ab}_{\mu}\partial_{\mu})c^{b}-\frac{\alpha}{2}b'^{a}b'^{a}+\frac{\alpha}{8}g^{2}f^{abc}f^{cde}\bar{c}^{e}\bar{c}^{a}c^{b}c^{d}\right]\,,
\label{cce6}
\end{equation}

\noindent and the equation of motion of $b'^a$ enforces the gauge condition (\ref{cce4}),

\begin{equation}
\frac{\delta S_{\mathrm{FP}}}{\delta b'^a}=\partial_{\mu}A^{a}_{\mu}-\alpha b'^a\,,
\label{cce7}
\end{equation}
The $SL(2,\mathbb{R})$ symmetry takes now the simpler form

\begin{eqnarray}
\delta\bar{c}^{a}&=&c^{a}\nonumber\\
\delta b'^a &=& 0\nonumber\\
\delta A^{a}_{\mu}&=&\delta c^{a}=0\,.
\label{cce8}
\end{eqnarray}
We see that the shift over the $b$ field simplifies the structure of the action and of the $SL(2,\mathbb{R})$ transformations. In particular, there are no  $(b'-c-{\bar c})$ interaction vertices. However, the use of the variable $b'$  introduces a more involved form for the BRST transformations,  eqs.\eqref{cce5}. In the following, we shall exploit the use of the shifted variable $b'$ whenever it will be more useful. 

\section{Construction of the Gribov-Zwanziger action}

In the last section we have established a connection between the Gribov problem in Curci-Ferrari and in linear covariant gauges. The latter were object of recent investigations in the context of the Refined Gribov-Zwanziger set up, see \cite{Capri:2015pja,Capri:2015ixa,Capri:2015nzw,Capri:2016aqq}. In particular, since the copies equation for Curci-Ferrari and linear covariant gauges is formally identical, the removal of Gribov copies in the Curci-Ferrari gauges follows exactly the same route as in linear covariant gauges. As a byproduct, the resulting Gribov-Zwanziger action in Curci-Ferrari gauges enjoys non-perturbative BRST invariance. From a different perspective, we can establish from the beginning a non-perturbative BRST quantization as already proposed in \cite{Capri:2015ixa,Capri:2015nzw}.  Following this prescription, we begin with the standard form of the gauge fixed action in the Curci-Ferrari gauges given by eq.(\ref{c1}) and employ the non-perturbative BRST quantization, namely

\begin{eqnarray}
S^{\mathrm{CF}}_{\mathrm{GZ}}&=& S_{\mathrm{YM}}+s_{\gamma^2}\int d^dx~\bar{c}^{a}\left[\partial_{\mu}A^{a}_{\mu}-\frac{\alpha}{2}\left(b^{h,a}-\frac{g}{2}f^{abc}\bar{c}^{b}c^{c}\right)\right]\nonumber\\
&+&\int d^dx\left(\bar{\varphi}^{ac}_{\mu}\left[\EuScript{M}(A^h)\right]^{ab}\varphi^{bc}_{\mu}-\bar{\omega}^{ac}_{\mu}\left[\EuScript{M}(A^h)\right]^{ab}\omega^{bc}_{\mu}+g\gamma^2f^{abc}A^{h,a}_{\mu}(\varphi+\bar{\varphi})^{bc}_{\mu}\right)\nonumber\\
&=& S_{\mathrm{YM}}+\int d^dx\left[b^{h,a}\partial_{\mu}A^{a}_{\mu}+\bar{c}^{a}\partial_{\mu}D^{ab}_{\mu}c^{b}-\frac{\alpha}{2}b^{h,a}b^{h,a}+\frac{\alpha}{2}gf^{abc}b^{h,a}\bar{c}^{b}c^{c}\right.\nonumber\\
&+&\left.\frac{\alpha}{8}g^{2}f^{abc}f^{cde}\bar{c}^{a}\bar{c}^{b}c^{d}c^{e}\right] 
+\int d^dx\left(\bar{\varphi}^{ac}_{\mu}\left[\EuScript{M}(A^h)\right]^{ab}\varphi^{bc}_{\mu}-\bar{\omega}^{ac}_{\mu}\left[\EuScript{M}(A^h)\right]^{ab}\omega^{bc}_{\mu}\right.\nonumber\\
&+&\left.g\gamma^2f^{abc}A^{h,a}_{\mu}(\varphi+\bar{\varphi})^{bc}_{\mu}\right)\,.
\label{gz1}
\end{eqnarray}
with $s_{\gamma^2}$ being  the non-perturbative and nilpotent BRST operator, see eq.\eqref{overview8.0}.

As already discussed in the context of linear covariant gauges, the proposed non-perturbative BRST quantization gives rise to a non-local action. From eq.\eqref{overview8.0}, even the non-perturbative BRST transformations are non-local. It is of uttermost interest to cast all the framework in a local fashion so that all the powerful machinery of local quantum field theories is  at our disposal. It turns out that it is possible to localize all this setting, as presented in \cite{Capri:2016aqq}. The extension to Curci-Ferrari gauges is straightforward and we will report the explicit local form in Sect.~7. However, before turning to this issue, we shall work out some features of the tree-level gluon propagator which do not require to go through all the localization procedure. 

For completeness, we present the form of the Gribov-Zwanziger action in Curci-Ferrari gauges in terms of the shifted field $b'^{h}$ \eqref{cce2},

\begin{eqnarray}
S^{\mathrm{CF}}_{\mathrm{GZ}}&=&S_{\mathrm{YM}}+\int d^dx\left[b'^{h,a}\partial_{\mu}A^{a}_{\mu}+\frac{1}{2}\bar{c}^{a}(\partial_{\mu}D^{ab}_{\mu}+D^{ab}_{\mu}\partial_{\mu})c^{b}-\frac{\alpha}{2}b'^{h,a}b'^{h,a}+\frac{\alpha}{8}g^{2}f^{abc}f^{cde}\bar{c}^{e}\bar{c}^{a}c^{b}c^{d}\right]\nonumber\\
&+&\int d^dx\left(\bar{\varphi}^{ac}_{\mu}\left[\EuScript{M}(A^h)\right]^{ab}\varphi^{bc}_{\mu}-\bar{\omega}^{ac}_{\mu}\left[\EuScript{M}(A^h)\right]^{ab}\omega^{bc}_{\mu}+g\gamma^2f^{abc}A^{h,a}_{\mu}(\varphi+\bar{\varphi})^{bc}_{\mu}\right)\,,\nonumber\\
\label{gz3}
\end{eqnarray}
which is invariant under the non-perturbative set of BRST transformations,

\begin{align}
s_{\gamma^2}A^{a}_{\mu}&=-D^{ab}_{\mu}c^b\,,     &&s_{\gamma^2}c^a=\frac{g}{2}f^{abc}c^bc^c\,, \nonumber\\
s_{\gamma^2}\bar{c}^a&=b'^{h,a}+\frac{g}{2}f^{abc}\bar{c}^bc^c\,,     &&s_{\gamma^2}b'^{h,a}=-\frac{g}{2}f^{abc}b'^{h,b}c^c+\frac{g^2}{8}f^{abc}f^{cde}\bar{c}^bc^dc^e\,, \nonumber\\
s_{\gamma^2}\varphi^{ab}_{\mu}&=\omega^{ab}_{\mu}\,,   &&s_{\gamma^2}\omega^{ab}_{\mu}=0\,, \nonumber\\
s_{\gamma^2}\bar{\omega}^{ab}_{\mu}&=\bar{\varphi}^{ab}_{\mu}-\gamma^2gf^{cdb}\int d^dy~A^{h,c}_{\mu}(y)\left[\EuScript{M}^{-1}(A^h)\right]^{da}_{yx}\,,         &&s_{\gamma^2}\bar{\varphi}^{ab}_{\mu}=0\,. 
\label{gz2}
\end{align}
As in the linear covariant gauges, the gap equation which determines the Gribov parameter reads

\begin{equation}
\frac{\partial\mathcal{E}_0}{\partial\gamma^2}=0\,\,\,\Rightarrow\,\,\,\langle gf^{abc}A^{h,a}_{\mu}(\varphi+\bar{\varphi})^{bc}_{\mu}\rangle=2d\gamma^2(N^2-1)\,. 
\label{gz23}
\end{equation}
The integration over $b'^h$ can be performed and the resulting action is

\begin{eqnarray}
S^{\mathrm{CF}}_{\mathrm{GZ}}&=&S_{\mathrm{YM}}+\int d^dx\left[\frac{(\partial_{\mu}A^{a}_{\mu})^{2}}{2\alpha}+\frac{1}{2}\bar{c}^{a}(\partial_{\mu}D^{ab}_{\mu}+D^{ab}_{\mu}\partial_{\mu})c^{b}+\frac{\alpha}{8}g^{2}f^{abc}f^{cde}\bar{c}^{e}\bar{c}^{a}c^{b}c^{d}\right]\nonumber\\
&+&\int d^dx\left(\bar{\varphi}^{ac}_{\mu}\left[\EuScript{M}(A^h)\right]^{ab}\varphi^{bc}_{\mu}-\bar{\omega}^{ac}_{\mu}\left[\EuScript{M}(A^h)\right]^{ab}\omega^{bc}_{\mu}+g\gamma^2f^{abc}A^{h,a}_{\mu}(\varphi+\bar{\varphi})^{bc}_{\mu}\right)\,.\nonumber\\
\label{gz5}
\end{eqnarray}
From the action \eqref{gz3} or, equivalently  \eqref{gz5}, the tree-level gluon propagator is given by 

\begin{equation}
\langle A^{a}_{\mu}(p)A^{b}_{\nu}(-p)\rangle=\delta^{ab}\left[\frac{p^2}{p^4+2g^2N\gamma^4}\left(\delta_{\mu\nu}-\frac{p_{\mu}p_{\nu}}{p^2}\right)+\frac{\alpha}{p^2}\frac{p_{\mu}p_{\nu}}{p^2}\right]\,.
\label{gz6}
\end{equation}
The transverse part turns out to be affected by the restriction of the domain of integration in the path integral due to the presence of $\gamma$, while the longitudinal part is equal to the perturbative result. We emphasize this is a tree-level computation only. The transverse part has the Gribov-type behavior. It is IR suppressed and its form factor goes to zero at zero-momentum. Also, this propagator violates positivity and as such, no physical particle interpretation can be attached to the gluon field. However, as already discussed in the Introduction, the Gribov-Zwanziger action suffers from IR instabilities and dimension-two condensates are formed. In the next section we take into account these effects and discuss their consequences for the gluon propagator.

\section{Dynamical generation of condensates}

\subsection{Refinement of the Gribov-Zwanziger action}

In the Landau gauge, it was noted that the Gribov-Zwanziger action suffers from IR instabilities, \cite{Dudal:2008sp}. In particular, already at the one-loop level it is possible to obtain a non-vanishing value for the dimension-two condensates which turn out to be proportional to the Gribov parameter $\gamma$. This shows that the formation of these condensates is deeply related to the presence of the Gribov horizon. In \cite{Capri:2015pja,Capri:2015nzw}, these results were extended to linear covariant gauges in the non-perturbative BRST framework. Also, analogous results were obtained for the maximal Abelian and the Coulomb gauges, \cite{Capri:2015pfa,Guimaraes:2015bra}. 

For the Curci-Ferrari gauges, we can proceed in full analogy with the case of the  linear covariant gauges. In particular, both condensates considered in \cite{Dudal:2008sp}, namely,

\begin{equation}
\langle A^{h,a}_{\mu}(x)A^{h,a}_{\mu}(x)\rangle\,\,\,\,\,\mathrm{and}\,\,\,\,\,\langle \bar{\varphi}^{ab}_{\mu}(x)\varphi^{ab}_{\mu}(x)-\bar{\omega}^{ab}_{\mu}(x)\omega^{ab}_{\mu}(x)\rangle\,,
\label{dgc5.0}
\end{equation}
are dynamically generated. This is easily proved by coupling  the aforementioned dimension-two operators to constant sources into the Gribov-Zwanziger action. Therefore, let us consider the generating functional $\mathcal{E}(m,J)$ defined as 

\begin{equation}
\mathrm{e}^{-V\mathcal{E}(m,J)}=\int\left[\EuScript{D}\Phi\right]\mathrm{e}^{-(S^{\mathrm{CF}}_{\mathrm{GZ}}+m\int d^dx~A^{h,a}_{\mu}A^{h,a}_{\mu}-J\int d^dx(\bar{\varphi}^{ab}_{\mu}\varphi^{ab}_{\mu}-\bar{\omega}^{ab}_{\mu}\omega^{ab}_{\mu}))}
\label{dgc6.0}
\end{equation}
with $m$ and $J$ being constant sources. Hence

\begin{eqnarray}
\langle \bar{\varphi}^{ab}_{\mu}\varphi^{ab}_{\mu}-\bar{\omega}^{ab}_{\mu}\omega^{ab}_{\mu} \rangle &=& -\frac{\partial \mathcal{E}(m,J)}{\partial J}\Big|_{m=J=0}\nonumber\\
\langle A^{h,a}_{\mu}A^{h,a}_{\mu}\rangle &=& \frac{\partial\mathcal{E}(m,J)}{\partial m}\Big|_{m=J=0}\,.
\label{dgc7.0}
\end{eqnarray}
At one-loop order, employing dimensional regularization,

\begin{equation}
{\cal E}(m,J)=\frac{(d-1)(N^2-1)}{2}\int \frac{d^dp}{(2\pi)^d}~\mathrm{ln}\left(p^2+\frac{2\gamma^4g^2N}{p^2+J}+2m\right)-d\gamma^4(N^2-1)\,,
\label{dgc8.0}
\end{equation}
which results in

\begin{equation}
\langle \bar{\varphi}^{ac}_{\mu}\varphi^{ac}_{\mu}-\bar{\omega}^{ac}_{\mu}\omega^{ac}_{\mu}\rangle = g^2\gamma^4N(N^2-1)(d-1)\int \frac{d^dp}{(2\pi)^d}\frac{1}{p^2}\frac{1}{(p^4+2g^2\gamma^4N)}
\label{dgc9.0}
\end{equation}
and

\begin{equation}
\langle A^{h,a}_{\mu}A^{h,a}_{\mu}\rangle = -2g^2\gamma^4N(N^2-1)(d-1)\int\frac{d^dp}{(2\pi)^d}\frac{1}{p^2}\frac{1}{(p^4+2g^2\gamma^4N)}\,.
\label{dgc10.0}
\end{equation}
From eqs.\eqref{dgc9.0},\eqref{dgc10.0}, we see immediately the presence of the Gribov parameter as a prefactor. This implies the non-triviality of the value of such condensates due to the restriction of the path integral domain to the Gribov region, encoded in $\gamma$. Also, as discussed in \cite{Capri:2015nzw,Dudal:2008xd}, the integrals appearing in \eqref{dgc9.0} and \eqref{dgc10.0} are perfectly convergent for $d=3,4$, while for $d=2$  develop an IR singularity. This behavior suggests the inclusion of \eqref{dgc5.0} to the Gribov-Zwanziger action for $d=3,4$, while keeping the action untouched for $d=2$. The absence of refinement of the Gribov-Zwanziger action in $d=2$ can be made more precise, see \cite{Capri:2015nzw,Dudal:2008xd}. Essentially, in $d=2$ it turns out to be impossible to remain within the Gribov region by introducing dimension two condensates \cite{Capri:2015nzw,Dudal:2008xd}. The same argument is easily extended to Curci-Ferrari gauges.

Taking into account these considerations, for the Refined Gribov-Zwanziger action in $d=3,4$ we obtain

\begin{eqnarray}
S^{\mathrm{CF}}_{\mathrm{RGZ}}&=& S_{\mathrm{YM}}+\int d^dx\left[b^{h,a}\partial_{\mu}A^{a}_{\mu}+\bar{c}^{a}\partial_{\mu}D^{ab}_{\mu}c^{b}-\frac{\alpha}{2}b^{h,a}b^{h,a}+\frac{\alpha}{2}gf^{abc}b^{h,a}\bar{c}^{b}c^{c}\right.\nonumber\\
&+&\left.\frac{\alpha}{8}g^{2}f^{abc}f^{cde}\bar{c}^{a}\bar{c}^{b}c^{d}c^{e}\right] 
+\int d^dx\left(\bar{\varphi}^{ac}_{\mu}\left[\EuScript{M}(A^h)\right]^{ab}\varphi^{bc}_{\mu}-\bar{\omega}^{ac}_{\mu}\left[\EuScript{M}(A^h)\right]^{ab}\omega^{bc}_{\mu}\right.\nonumber\\
&+&\left.g\gamma^2f^{abc}A^{h,a}_{\mu}(\varphi+\bar{\varphi})^{bc}_{\mu}\right)+\frac{m^2}{2}\int d^dx\,A^{h,a}_{\mu}A^{h,a}_{\mu}-M^2\int d^dx\,\left(\bar{\varphi}^{ab}_{\mu}\varphi^{ab}_{\mu}-\bar{\omega}^{ab}_{\mu}\omega^{ab}_{\mu}\right)\,.
\label{dgc11.0}
\end{eqnarray}
while in $d=2$ the Gribov-Zwanziger action is left unmodified and expression \eqref{gz1} is preserved.

\subsection{A remark on the gluon-ghost condensate}

In the last decade, much effort has been undertaken to understand the QCD vacuum and, in particular, the pure Yang-Mills vacuum. Much attention was devoted to the dynamical formation of condensates which could introduce non-perturbative effects related to chiral symmetry breaking (in the specific case of QCD) and color confinement. Also, dimension-two gluon condensates were on the mainstream of analytical and numerical approaches to confinement due to the possibility of giving rise to a possible mechanism for dynamical mass generation. On the other hand, the dimension-two gluon condensate $\langle A^{a}_{\mu}A^{a}_{\mu}\rangle$ is not gauge invariant for a generic choice of a covariant renormalizable gauge and a direct physical interpretation is unclear. Moreover, a genuine gauge invariant expression  is provided by $\langle A^h_\mu A^h_\mu\rangle$. Albeit gauge invariant, this quantity  is highly non-local, with the notable exception of the  Landau gauge, where $\mathcal{A}^{2}_{\mathrm{min}}$ reduces to the simple expression $A^{a}_{\mu}A^{a}_{\mu}$. This is a very special feature of Landau gauge. On the other hand, the existence of other dimension-two condensates is also possible. A particular example is the ghost condensate $\langle \bar{c}^ac^a\rangle$. Though, Yang-Mills theories quantized in Landau gauge displays an additional  Ward identity, the anti-ghost equation of motion, which forbids the existence of $\langle \bar{c}^ac^a\rangle$. The same Ward identity holds for linear covariant gauges. Therefore, in these cases, just the gluon condensate is allowed. However, in the Curci-Ferrari gauges, the anti-ghost equation is not a Ward identity anymore and there is no a priori reason to exclude the condensate $\langle \bar{c}^ac^a\rangle$. Hence, we can introduce the general term

\begin{equation}
\tilde{S}_{\mathrm{cond}}=\int d^dx\left(\kappa_1 A^{a}_{\mu}A^{a}_{\mu}+\kappa_2\bar{c}^ac^a\right)\,,
\label{dgc2.0}
\end{equation}
and demand invariance under BRST and the $SL(2,\mathbb{R})$ symmetry. The latter does not impose any constraint on the coefficients $\kappa_1$ and $\kappa_2$. BRST, however, does\footnote{There is no difference in making use of the standard BRST or the non-perturbative one, due to the fact that for $(A,c,\bar{c})$ these transformations are identical.}:

\begin{equation}
s\tilde{S}_{\mathrm{cond}} = \int d^dx\left(2\kappa_1(\partial_{\mu}A^{a}_{\mu})c^a+k_2b'ac^a\right)\approx 0\,\,\,\,\Rightarrow\,\,\,\, \kappa_2=-2\alpha\kappa_1\,,
\label{dgc3.0}
\end{equation}
where the symbol $\approx$ denotes that we have used the equations of motion. Therefore, modulo a prefactor, the (on-shell) BRST invariant operator is

\begin{equation}
\EuScript{O}=\frac{1}{2}A^{a}_{\mu}A^{a}_{\mu}-\alpha\bar{c}^ac^a\,.
\label{dgc4.0}
\end{equation}
Some remarks concerning expression \eqref{dgc4.0} are in order: $(i)$ The limit $\alpha\rightarrow 0$ corresponds to the Landau gauge. In this case, the operator \eqref{dgc4.0} reduces to the dimension-two gluon operator $A^{a}_{\mu}A^{a}_{\mu}$ and no ghost condensate is included. $(ii)$ As is well-known, the presence of the quartic interaction term of Faddeev-Popov ghosts is  responsible for (eventually) generating a non-vanishing ghost condensate $\langle \bar{c}^ac^a\rangle$. 

Evidences for the existence of the condensate \eqref{dgc4.0} were presented in \cite{Kondo:2001nq,Dudal:2003gu}. In \cite{Kondo:2001nq} the modification of the OPE for the gluon and ghost due to the dimension two-condensate \eqref{dgc4.0} was pointed out,  while in \cite{Dudal:2003gu} an effective potential analysis was carried out. Unfortunately, the lack of lattice simulations results for Curci-Ferrari gauges does not allow us to give more conclusive statements concerning the relevance of the condensate \eqref{dgc4.0}. 

Nevertheless, within the new non-perturbative BRST framework, we introduced directly the gauge invariant quantity $\langle A^h_{\mu}A^h_{\mu}\rangle$ in the refinement of the Gribov-Zwanziger action. This condensate, as the gluon-ghost condensate (\ref{dgc4.0}), reduces to $\langle A^a_{\mu}A^{a}_{\mu}\rangle$ in the Landau gauge. In this sense, the introduction of both condensates seems to be redundant. Moreover, as will be discussed in Sect.~7, we have a local set up for $\langle A^h_{\mu}A^h_{\mu}\rangle$, evading the main difficulties that earlier studies had to deal with this operator. In summary, $\langle A^h_{\mu}A^h_{\mu}\rangle$ should be responsible to carry all physical information of \eqref{dgc4.0}. A very attractive feature is that the gauge invariance of $\langle A^h_{\mu}A^h_{\mu}\rangle$ together with the non-perturbative BRST symmetry gives to us full control of the independence from $\alpha$ of correlation funtions of gauge invariant operators. Therefore, the inclusion of \eqref{dgc4.0} seems to be superfluous, due to the use of the operator $A^h_\mu A^h_\mu$. 

We remark that the formation of different ghost condensates was also studied in Curci-Ferrari gauges, see \cite{Dudal:2002ye,Lemes:2002jv}. In principle, we should take them into account as well. However, in this work we are concerned with the behavior of the gluon propagator and, for this purpose, the inclusion of these extra condensates is irrelevant. Moreover, these condensates affect the ghost propagator and, again, it  would be  desirable to have access to lattice simulations for such propagator in order to estimate the relevance played by these novel condensates.

\section{Gluon propagator}

In the last section we discussed non-trivial dynamical effects generated in Curci-Ferrari gauges. As it happens in the Gribov-Zwanziger theory in the gauges already studied in the literature, the presence of the Gribov horizon contributes to the formation of dimension-two condensates. The Refined Gribov-Zwanziger action in Curci-Ferrari gauges is given by \eqref{dgc11.0}, where such condensates are taken into account from the beginning through the presence of the dynamical parameters $(M^2, m^2)$. Hence, we can easily compute the gluon propagator out of \eqref{dgc11.0}, namely

\begin{equation}
\langle A^{a}_{\mu}(p)A^{b}_{\nu}(-p)\rangle_{d=3,4}=\delta^{ab}\left[\frac{p^2+M^2}{(p^2+m^2)(p^2+M^2)+2g^2\gamma^4N}\left(\delta_{\mu\nu}-\frac{p_{\mu}p_{\nu}}{p^2}\right)+\frac{\alpha}{p^2}\frac{p_{\mu}p_{\nu}}{p^2}\right]\,,
\label{cfgp1.0}
\end{equation}
while in $d=2$, we use the Gribov-Zwanziger action \eqref{gz1},

\begin{equation}
\langle A^{a}_{\mu}(p)A^{b}_{\nu}(-p)\rangle_{d=2}=\delta^{ab}\left[\frac{p^2}{p^4+2g^2\gamma^4N}\left(\delta_{\mu\nu}-\frac{p_{\mu}p_{\nu}}{p^2}\right)+\frac{\alpha}{p^2}\frac{p_{\mu}p_{\nu}}{p^2}\right]\,.
\label{cfgp2.0}
\end{equation}
Several remarks are in order.  For $d=3,4$,

\begin{itemize}

\item The form factor of the transverse part of the propagator is IR suppressed, positivity violating and attains a finite non-vanishing value at zero-momentum, a property which follows from the inclusion of the dimension two condensate of the auxiliary fields $\langle \bar{\varphi}\varphi-\bar{\omega}\omega\rangle$. Also, at tree-level, this form factor is independent from  $\alpha$. Hence, the transverse component of the gluon propagator displays the so-called decoupling/massive behavior.

\item The limit $\alpha\rightarrow 0$ brings us back to the gluon propagator for the Refined Gribov-Zwanziger action in the Landau gauge.

\item In the linear covariant gauges, the longitudinal part of the gluon propagator does not receive non-perturbative corrections. It remains as in perturbation theory, which is known to be just the tree-level result without quantum corrections. However, in Curci-Ferrari gauges, non-linearity jeopardizes this property as follows, for example, from the existence of the interaction vertex $b$-$c$-${\bar c}$. Therefore, inhere we expect that loop corrections will affect the longitudinal sector, although an explicit verification is far beyond the scope of this work. 

\end{itemize}

\noindent In the case of $d=2$,

\begin{itemize}

\item Since in $d=2$ the Gribov-Zwanziger action does not suffer from refinement, the gluon propagator is of the Gribov-type \textit{i.e.} the transverse part is IR suppressed, positivity violating and vanishes at zero-momentum. This characterizes the so-called  scaling behavior.

\item As in $d=3,4$, the Landau propagator is easily obtained for $\alpha\rightarrow 0$, giving the scaling Gribov gluon propagator in $d=2$.

\end{itemize}

\noindent From these comments we can conclude that, for $d=3,4$, the transverse gluon propagator displays a decoupling/massive behavior while in $d=2$ it is of scaling type. This is precisely the same behavior obtained in the Landau gauge and reported by very large lattice simulations. As pointed out in \cite{Capri:2015nzw,Capri:2015pfa,Guimaraes:2015bra}, this feature is more general than a particular property of  the Landau gauge, being also present in the linear covariant, maximal Abelian and Coulomb gauges. Inhere, we provide evidence that this property should also hold in Curci-Ferrari gauges. The novelty here with respect to the gauges already studied is the non-triviality of the longitudinal part which, due to the very non-linear character of the Curci-Ferrari gauges,  might very well acquire corrections from higher loops. 

\section{Local Refined Gribov-Zwanziger action in Curci-Ferrari gauges}

In this section we present a localization procedure to cast the action \eqref{gz1} and the transformations \eqref{gz2} in a suitable local fashion. This puts the (Refined) Gribov-Zwanziger action in Curci-Ferrari gauges within the  well-developed realm of local quantum field theory. Before starting the description of the procedure, we emphasize the already mentioned feature that the original formulation of Gribov-Zwanziger action in the Landau gauge \eqref{intro5.0} relies on the introduction of a non-local horizon function, displaying thus a non-local character. As shown previously, this non-locality can be handled through the introduction of suitable auxiliary fields which provide a local and renormalizable framework. 

Nevertheless, as soon as we introduce the gauge invariant field $A^h$, we introduce a new source of non-locality, see eq.\eqref{overview1.0}. Hence, even after the introduction of the auxiliary fields introduced in the standard construction, the resulting action is still non-local due to the explicit presence of $A^h$. 

The localization of the transverse gauge invariant field $A^h$ is performed by the introduction of a Stueckelberg-type field $\xi^a$ in the form

\begin{equation}
h=\mathrm{e}^{ig\xi^aT^a}\,.
\label{lcf1.0}
\end{equation}
With \eqref{lcf1.0}, we rewrite the $A^h$ field as

\begin{equation}
A^h_{\mu}=h^{\dagger}A_{\mu}h+\frac{i}{g}h^{\dagger}\partial_{\mu}h\,,
\label{lcf2.0}
\end{equation}
where a matrix notation has been  employed. Expression \eqref{lcf2.0} is local albeit non-polynomial. For a $SU(N)$ element $v$, $A^h$ is left invariant under the gauge transformations

\begin{equation}
A'_{\mu}=v^{\dagger}A_{\mu}v+\frac{i}{g}v^{\dagger}\partial_{\mu}v\,,\,\,\,\,\, h'=v^{\dagger}h\,\,\,\,\, \mathrm{and}\,\,\,\,\, h'^{\dagger}=h^{\dagger}v\,,
\label{lcf3.0}
\end{equation}
{\it i.e.}
\begin{equation} 
 ({A^{h}_\mu})'  \leftarrow A^h_{\mu}  \;. \label{gah}
\end{equation}
Although gauge invariance of $A^h$ is guaranteed by \eqref{lcf3.0}, we still have to impose the transversality condition of $A^h$. This is done by means of a Lagrange multiplier $\tau^a$ which enforces this constraint, namely, we introduce the following term

\begin{equation}
S_{\tau}=\int d^dx\,\tau^a\partial_{\mu}A^{h,a}_{\mu}\,.
\label{lcf4.0}
\end{equation}
Solving the transversality condition $\partial A^h=0$ for $\xi$, we obtain the non-local expression \eqref{overview1.0} for $A^h$, see, for example, Appendix A of \cite{Capri:2015ixa}. Then, the Gribov-Zwanziger action in Curci-Ferrari gauges can be expressed in local form as follows,

\begin{eqnarray}
S^{\mathrm{loc}}_{\mathrm{CF}} &=&  S_{\mathrm{YM}} + \int d^dx\left(b^{h,a}\partial_{\mu}A^{a}_{\mu}-\frac{\alpha}{2}b^{h,a}b^{h,a}+\bar{c}^{a}\partial_{\mu}D^{ab}_{\mu}c^{b}+\frac{\alpha}{2}gf^{abc}b^{a}\bar{c}^{b}c^{c}\right.\nonumber\\
&+&\left.\frac{\alpha}{8}g^{2}f^{abc}f^{cde}\bar{c}^{a}\bar{c}^{b}c^{d}c^{e}\right)-\int d^dx\left(\bar{\varphi}^{ac}_{\mu}\left[\EuScript{M}(A^h)\right]^{ab}\varphi^{bc}_{\mu}-\bar{\omega}^{ac}_{\mu}\left[\EuScript{M}(A^h)\right]^{ab}\omega^{bc}_{\mu}\right.\nonumber\\
&+&\left.g\gamma^2f^{abc}A^{h,a}_{\mu}(\varphi+\bar{\varphi})^{bc}_{\mu}\right)+\int d^dx~\tau^a\partial_{\mu}A^{h,a}_{\mu}\,,\nonumber\\
\label{lcf5.0}
\end{eqnarray} 
with $A^h$ given by \eqref{lcf3.0}. 

The non-perturbative BRST transformations, which correspond to a symmetry of \eqref{lcf5.0}, are also non-local. As shown in \cite{Capri:2016aqq}, the localization of these transformations is achieved through the introduction of extra auxiliary fields. Before doing this, we note that the BRST transformations for $\tau$ and $\xi$ (written implicitly in terms of $h$) are

\begin{equation}
sh=-igch\,\,\,\,\,\,\,\,\mathrm{and}\,\,\,\,\,\,\,\,s\tau^a=0\,.
\label{lcf6.0}
\end{equation}
Proceeding with the localization of the non-perturbative BRST transformations, we make use of the following trick: We rewrite the horizon function $H(A^h)$ in the path integral  as

\begin{equation}
\mathrm{e}^{-\gamma^4 H(A^{h})}=\mathrm{e}^{-\frac{\gamma^4}{2} H(A^{h})}\mathrm{e}^{-\frac{\gamma^4}{2} H(A^{h})}\,.
\label{lcf7.0}
\end{equation}
Now, employing the same localization procedure used in the standard Gribov-Zwanziger framework, we obtain

\begin{equation}
\mathrm{e}^{-\frac{\gamma^4}{2} H(A^{h})}=\int \left[\EuScript{D}\varphi\right]\left[\EuScript{D}\bar{\varphi}\right]\left[\EuScript{D}\omega\right]\left[\EuScript{D}\bar{\omega}\right]\mathrm{e}^{-\int d^dx\left(-\bar{\varphi}^{ac}_{\mu}\EuScript{M}^{ab}(A^h)\varphi^{bc}_{\mu}+\bar{\omega}^{ac}_{\mu}\EuScript{M}^{ab}(A^h)\omega^{bc}_{\mu}+g\frac{\gamma^2}{\sqrt{2}}f^{abc}A^{h,a}_{\mu}(\varphi+\bar{\varphi})^{bc}_{\mu}\right)}\,,
\label{lcf8.0}
\end{equation}
and

\begin{equation}
\mathrm{e}^{-\frac{\gamma^4}{2} H(A^{h})}=\int \left[\EuScript{D}\beta\right]\left[\EuScript{D}\bar{\beta}\right]\left[\EuScript{D}\zeta\right]\left[\EuScript{D}\bar{\zeta}\right]\mathrm{e}^{-\int d^dx\left(-\bar{\beta}^{ac}_{\mu}\EuScript{M}^{ab}(A^h)\beta^{bc}_{\mu}+\bar{\zeta}^{ac}_{\mu}\EuScript{M}^{ab}(A^h)\zeta^{bc}_{\mu}-g\frac{\gamma^2}{\sqrt{2}}f^{abc}A^{h,a}_{\mu}(\beta+\bar{\beta})^{bc}_{\mu}\right)}\,.
\label{lcf9.0}
\end{equation}
In \eqref{lcf8.0}, the fields $(\varphi,\bar{\varphi},\omega,\bar{\omega})$ are Zwanziger's localizing fields, $(\beta,\bar{\beta})$ are commuting ones while $(\zeta,\bar{\zeta})$ are anti-commuting and play the same role as Zwanziger's fields. The resulting Gribov-Zwanziger action is given by

\begin{eqnarray}
S^{\mathrm{loc}}_{{\mathrm{CF}}}&=& S_{\mathrm{YM}}+\int d^dx\left(b^{h,a}\partial_{\mu}A^{a}_{\mu}-\frac{\alpha}{2}b^{h,a}b^{h,a}+\bar{c}^{a}\partial_{\mu}D^{ab}_{\mu}c^{b}+\frac{\alpha}{2}gf^{abc}b^{a}\bar{c}^{b}c^{c}\right.\nonumber\\
&+&\left.\frac{\alpha}{8}g^{2}f^{abc}f^{cde}\bar{c}^{a}\bar{c}^{b}c^{d}c^{e}\right)-\int d^dx\left(\bar{\varphi}^{ac}_{\mu}\EuScript{M}^{ab}(A^h)\varphi^{bc}_{\mu}-\bar{\omega}^{ac}_{\mu}\EuScript{M}^{ab}(A^h)\omega^{bc}_{\mu}\right.\nonumber\\
&-&\left.g\frac{\gamma^2}{\sqrt{2}}f^{abc}A^{h,a}_{\mu}(\varphi+\bar{\varphi})^{bc}_{\mu}\right)-\int d^dx\left(\bar{\beta}^{ac}_{\mu}\EuScript{M}^{ab}(A^h)\beta^{bc}_{\mu}-\bar{\zeta}^{ac}_{\mu}\EuScript{M}^{ab}(A^h)\zeta^{bc}_{\mu}\right.\nonumber\\
&+&\left.g\frac{\gamma^2}{\sqrt{2}}f^{abc}A^{h,a}_{\mu}(\beta+\bar{\beta})^{bc}_{\mu}\right)+\int d^dx~\tau^a\partial_{\mu}A^{h,a}_{\mu}\,. \nonumber\\
\label{lcf10.0}
\end{eqnarray}
The local Gribov-Zwanziger action written as \eqref{lcf10.0} is invariant under the following local non-perturbative BRST transformations,

\begin{align}
s_{l}A^{a}_{\mu}&=-D^{ab}_{\mu}c^{b}\,,     && s_{l}\varphi^{ab}_{\mu}=\omega^{ab}_{\mu}\,,&&&      s_{l}h =-igch\,,&&&& s_{l}\beta^{ab}_{\mu}=\omega^{ab}_{\mu}\,,\nonumber\\
s_{l}c^a&=\frac{g}{2}f^{abc}c^{b}c^{c}\,,     && s_{l}\omega^{ab}_{\mu}=0\,,&&&     s_{l}A^{h,a}_{\mu}=0\,,&&&& s_{l}\bar{\zeta}^{ab}_{\mu}=0\,,\nonumber\\
s_{l}\bar{c}^{a}&=b^{h,a}\,,   && s_{l}\bar{\omega}^{ab}_{\mu}=\bar{\varphi}^{ab}_{\mu}+\bar{\beta}^{ab}_{\mu}\,,&&&     s_{l}\tau^{a} =0\,,&&&&s_{l}\zeta^{ab}_{\mu}=0\,.\nonumber\\
s_{l}b^{h,a}&=0\,,         && s_{l}\bar{\varphi}^{ab}_{\mu}=0\,,&&&s_{l}\bar{\beta}^{ab}_{\mu}=0\,,
\label{lcf11.0}
\end{align}
It is an immediate check that $s_l$ is nilpotent, $s^2_l=0$. Integration over $(\beta,\bar{\beta},\zeta,\bar{\zeta})$ gives back the non-local BRST transformations \eqref{gz2}.

In local fashion, the refinement of the Gribov-Zwanziger action is obtained by the introduction of the following term to \eqref{lcf10.0},

\begin{equation}
S_{\mathrm{cond}}=\int d^dx\left[\frac{m^2}{2}A^{h,a}_{\mu}A^{h,a}_{\mu}+M^2\left(\bar{\omega}^{ab}_{\mu}\omega^{ab}_{\mu}-\bar{\varphi}^{ab}_{\mu}\varphi^{ab}_{\mu}-\bar{\beta}^{ab}_{\mu}\beta^{ab}_{\mu}+\bar{\zeta}^{ab}_{\mu}\zeta^{ab}_{\mu}\right)\right]\,.
\label{lcf12.0}
\end{equation}
The resulting Refined Gribov-Zwanziger action, written in local form and invariant under \eqref{lcf11.0} is

\begin{eqnarray}
S^{\mathrm{RGZ}}_{\mathrm{CF}}&=&S_{\mathrm{YM}}+\int d^dx\left(b^{h,a}\partial_{\mu}A^{a}_{\mu}-\frac{\alpha}{2}b^{h,a}b^{h,a}+\bar{c}^{a}\partial_{\mu}D^{ab}_{\mu}c^{b}+\frac{\alpha}{2}gf^{abc}b^{a}\bar{c}^{b}c^{c}\right.\nonumber\\
&+&\left.\frac{\alpha}{8}g^{2}f^{abc}f^{cde}\bar{c}^{a}\bar{c}^{b}c^{d}c^{e}\right)-\int d^dx\left(\bar{\varphi}^{ac}_{\mu}\EuScript{M}^{ab}(A^h)\varphi^{bc}_{\mu}-\bar{\omega}^{ac}_{\mu}\EuScript{M}^{ab}(A^h)\omega^{bc}_{\mu}\right.\nonumber\\
&-&\left.g\frac{\gamma^2}{\sqrt{2}}f^{abc}A^{h,a}_{\mu}(\varphi+\bar{\varphi})^{bc}_{\mu}\right)-\int d^dx\left(\bar{\beta}^{ac}_{\mu}\EuScript{M}^{ab}(A^h)\beta^{bc}_{\mu}-\bar{\zeta}^{ac}_{\mu}\EuScript{M}^{ab}(A^h)\zeta^{bc}_{\mu}\right.\nonumber\\
&+&\left.g\frac{\gamma^2}{\sqrt{2}}f^{abc}A^{h,a}_{\mu}(\beta+\bar{\beta})^{bc}_{\mu}\right)+\int d^dx~\tau^a\partial_{\mu}A^{h,a}_{\mu}+\int d^dx\left[\frac{m^2}{2}A^{h,a}_{\mu}A^{h,a}_{\mu}
\right.\nonumber\\
&+&\left.M^2\left(\bar{\omega}^{ab}_{\mu}\omega^{ab}_{\mu}-\bar{\varphi}^{ab}_{\mu}\varphi^{ab}_{\mu}-\bar{\beta}^{ab}_{\mu}\beta^{ab}_{\mu}+\bar{\zeta}^{ab}_{\mu}\zeta^{ab}_{\mu}\right)\right]\,. 
\label{lcf14.0}
\end{eqnarray}
The Refined Gribov-Zwanziger action \eqref{lcf14.0} is an effective action which takes into account the presence of Gribov copies in the standard Faddeev-Popov procedure in Curci-Ferrari gauges. Moreover, this action also incorporates further non-perturbative dynamics effects as the formation of dimension-two condensates. All this setting is written in local fashion and enjoys non-perturbative BRST symmetry \eqref{lcf11.0} which ensures gauge parameter independence of correlation functions of gauge invariant composite operators, see \cite{Capri:2016aqq} for a purely  algebraic proof of this statement.

For completeness, we exhibit the Slavnov-Taylor identity associated with the local non-perturbative invariance of the Refined Gribov-Zwanziger action in Curci-Ferrari gauges. To do so, we introduce the following source action to control the non-linearity of the non-perturbative BRST transformations,

\begin{equation}
S_{\mathrm{sources}}=\int d^dx\left(-\Omega^{a}_{\mu}D^{ab}_{\mu}c^b+\frac{g}{2}f^{abc}L^a c^b c^c+K^a s_l\xi^a\right)\,,
\label{lcf15.0}
\end{equation}
with $s_l\Omega^{a}_{\mu}=s_l L^{a}=s_l K^a=0$. The resulting action $\Sigma$ defined as

\begin{equation}
\Sigma = S^{\mathrm{RGZ}}_{\mathrm{CF}} + S_{\mathrm{sources}}\,,
\label{lcf16.0}
\end{equation}
satisfies the following Slavnov-Taylor identity,

\begin{eqnarray}
\mathcal{S}(\Sigma)&=&\int d^dx\left(\frac{\delta\Sigma}{\delta A^a_\mu}\frac{\delta\Sigma}{\delta\Omega^a_{\mu}}+\frac{\delta\Sigma}{\delta c^a}\frac{\delta\Sigma}{\delta L^a}+\frac{\delta\Sigma}{\delta\xi^a}\frac{\delta\Sigma}{\delta K^a}+b^{h,a}\frac{\delta\Sigma}{\delta \bar{c}^a}+\omega^{ab}_{\mu}\frac{\delta\Sigma}{\delta\varphi^{ab}_{\mu}}+(\bar{\varphi}+\bar{\beta})^{ab}_{\mu}\frac{\delta\Sigma}{\delta\bar{\omega}^{ab}_{\mu}}\right.\nonumber\\
&+&\left.\omega^{ab}_{\mu}\frac{\delta\Sigma}{\delta\beta^{ab}_{\mu}}\right)=0\,.
\label{lcf17.0}
\end{eqnarray}
As is well known, the external sources $(\Omega^a_\mu, L^a, K^a)$ play exactly the same role of the so-called anti-fields of the Batalin-Vilkovisky formalism. As shown recently in the case of the linear covariant gauges \cite{Capri:2016aqq}, the Slavnov-Taylor identity \eqref{lcf16.0} can be employed to extract non-perturbative properties related to the physical content of the theory. In particular, the construction outlined in details in  \cite{Capri:2016aqq}, see sections III and IV, can be immediately repeated in the present case. This leads to show that the correlation functions of composite operators belonging to the cohomology of the non-perturbative BRST nilpotent operator turn out to be independent from the gauge parameter entering the gauge fixing condition. Further, examples of non-trivial composite operators exhibiting  a K\"all\'en-Lehmann representation with a positive spectral density can be worked out by means of the use of the $i$-particle formalism developed in \cite{Baulieu:2009ha}. Due to the positiveness of the spectral density, these operators can be directly linked with the physical spectrum of a confining Yang-Mills theory, as shown in the case of the glueball states \cite{Dudal:2010cd,Dudal:2013wja}. 

\section{Conclusions}

In this paper we have addressed the issue of  the quantization of Yang-Mills theories in a class of non-linear gauges, the Curci-Ferrari gauges, by taking into account the existence of Gribov copies. By exploiting the formal similarity with the Gribov problem in linear covariant gauges, the recently non-perturbative BRST transformations introduced in \cite{Capri:2015ixa} have been  used to achieve a non-perturbative BRST quantization scheme in the Curci-Ferrari gauges, resulting in an action akin to the Gribov-Zwanziger action in linear covariant gauges \cite{Capri:2015ixa,Capri:2015nzw}.

As is known, the Gribov problem entails modifications on the IR behavior of  the theory due to its non-perturbative nature. The so called Gribov-Zwanziger framework enables us to take into account the effects of the Gribov copies within the realm of a local Euclidean quantum field theory. Further, taking into account the dynamical generation of dimension two condensates, gives rise to the Refined Gribov-Zwanziger framework. As discussed previously, the introduction of such novel effects is not consistent in $d=2$. An immediate consequence of this fact is the difference of the gluon propagator behavior in different dimensions: in $d=3,4$, the dimension-two condensate of auxiliary Zwanziger's field yields  a decoupling/massive behavior for the transverse part of the propagator, while in $d=2$ this condensate cannot be consistently introduced and the transverse component is of the scaling-type. Remarkably, this  different behavior of the transverse component of the gluon propagator  has been  also observed in other  gauges, namely:  Landau, linear covariant, maximal Abelian and Coulomb gauges, see \cite{Capri:2015nzw,Dudal:2008xd,Capri:2015pfa,Guimaraes:2015bra}. It strongly suggests a kind of universal behavior for the transverse component of the gluon propagator,  as far as space-time dimensions are concerned. 

Nevertheless, unlike the case of the Landau and linear covariant gauges, the non-linearity of the Curci-Ferrari gauges might introduce  non-trivial effects in the longitudinal sector which  cannot be anymore protected from higher loops corrections. 

The construction of the non-perturbative BRST invariant (Refined) Gribov-Zwanziger action in Curci-Ferrari gauges relies on the use of the non-local variable $A^h$. In this work we provided a procedure which allows the localization of the action as well as of the non-perturbative BRST transformations. The resulting local Refined Gribov-Zwanziger action in Curci-Ferrari gauges provides then a suitable arena to apply standard local quantum field theories techniques which might  open new future investigations such as:  $i)$ study of the all order renormalizability of the action \eqref{gz1}, $ii)$ a better understanding of the longitudinal sector of the gluon propagator when higher orders effect are taken into account,  $iii)$ a detailed investigation of the ghost two-point function and its possible relationship with the ghost condensates already observed in the Curci-Ferrari gauges \cite{Lemes:2002jv}. Although this gauge is not yet well exploited from the  lattice point of view,  we hope that our results will stimulate future investigations in this direction, providing then an interesting interplay between  analytical and numerical results.

\section*{Acknowledgements}
The Conselho Nacional de Desenvolvimento Cient\'{i}fico e Tecnol\'{o}gico (CNPq-Brazil), The Coordena\c c\~ao de Aperfei\c coamento de Pessoal de N\'ivel Superior (CAPES), the Pr\'o-Reitoria de Pesquisa, P\'os-Gradua\c c\~ao e Inova\c c\~ao (PROPPI-UFF) and the DAAD are acknowledged for financial support. A.D.P. would like to express his gratitude to the Albert Einstein Institute for hospitality and support.

\appendix

\section{On the $SL(2,\mathbb{R})$ symmetry}

As pointed out in Sect.~3, the symmetry generated by $\delta$ is crucial for the proof of the perturbative renormalizability of Yang-Mills theory quantized in Curci-Ferrari gauges. This symmetry is part of a $SL(2,\mathbb{R})$ invariance. In this appendix, we exhibit the full Nakanishi-Ojima algebra, which contains the $SL(2,\mathbb{R})$ algebra as a subalgebra.  We refer the reader to \cite{Delduc:1989uc,Dudal:2002ye} for further details. 

The nilpotent perturbative BRST operator $s$ acts on the fields $(A,\bar{c},c,b)$ as in eq.(29). Nevertheless, it is possible to define the anti-BRST transformations generated by the nilpotent operator $\bar{s}$ as

\begin{eqnarray}
\bar{s}A^{a}_{\mu}&=&-D^{ab}_{\mu}\bar{c}^b\nonumber\\
\bar{s}c^a&=&-b^a+gf^{abc}c^{b}\bar{c}^c\nonumber\\
\bar{s}\bar{c}^a&=&\frac{g}{2}f^{abc}\bar{c}^b\bar{c}^c\nonumber\\
\bar{s}b^a&=&-gf^{abc}b^b\bar{c}^c\,.
\label{ap1}
\end{eqnarray} 
Likewise, the operator $\delta$ acts as in eq.(30) and we can define an operator $\bar{\delta}$ which acts on the $(A,\bar{c},c,b)$ as

\begin{eqnarray}
\bar{\delta}c^a&=&\bar{c}^a\nonumber\\
\bar{\delta}b^a&=&\frac{g}{2}f^{abc}\bar{c}^b\bar{c}^c\nonumber\\
\bar{\delta}A^{a}_{\mu}&=&\bar{\delta}\bar{c}^a=0\,.
\label{ap2}
\end{eqnarray}
The operators $s$, $\bar{s}$, $\delta$, $\bar{\delta}$ with the addition of the Faddeev-Popov ghost number operator $\delta_{\mathrm{FP}}$ form the so-called Nakanishi-Ojima algebra, given by

\begin{align}
\left\{s,\bar{s}\right\}&=0\,,     &&[\delta,\bar{\delta}]=\delta_{\mathrm{FP}}\,, \nonumber\\
[\delta,\delta_{\mathrm{FP}}]&=-2\delta\,,     &&[\bar{\delta},\delta_{\mathrm{FP}}]=2\bar{\delta}\,, \nonumber\\
[s,\delta_{\mathrm{FP}}]&=-s\,,   &&[\bar{s},\delta_{\mathrm{FP}}]=\bar{s}\,, \nonumber\\
[s,\delta]&=0\,,         &&[\bar{s},\bar{\delta}]=0\,, \nonumber\\
[s,\bar{\delta}]&=-\bar{s}\,,      &&[\bar{s},\delta]=-s\,. 
\label{ap3}
\end{align}
From \eqref{ap3}, it is clear that the operators $\delta$, $\bar{\delta}$ and $\delta_{\mathrm{FP}}$ generate a $SL(2,\mathbb{R})$ algebra. For this reason, the symmetry generated by $\delta$, defined by eq.(30) is known as the $SL(2,\mathbb{R})$ symmetry, although one should keep in mind the existence of the aforementioned algebraic structure. 


\end{document}